\documentclass[runningheads]{llncs}
\usepackage[T1]{fontenc}
\usepackage{graphicx}
\usepackage{multirow}
\usepackage{booktabs}

\usepackage[colorinlistoftodos]{todonotes}

\begin{document}
\title{Where is Dmitry going? Framing `migratory'
decisions in the criminal underground}
\titlerunning{Where is Dmitry going?}
%
\author{Luca Allodi \and
Roy Ricaldi\and
Jai Wientjes\and
Adriana Radu}
\authorrunning{L. Allodi et al.}
%
\institute{Eindhoven University of Technology, NL
\url{https://security1.win.tue.nl}}
\maketitle              
%

%
%
%
\section{Introduction}

The cybercriminal underground consists of hundreds of forum communities that function as marketplaces and information-exchange platforms for both established and wannabe cybercriminals. The ecosystem is continuously evolving, with users `migrating’ between forums and platforms. The emergence of cybercrime communities in Telegram and Discord only highlights the rising fragmentation and adaptability of the ecosystem. 
In this position paper, we explore the economic incentives and trust-building mechanisms that may drive a participant (hereafter, `\textit{Dmitry')} of the cybercriminal underground ecosystem to `migrate' from one forum or platform to another. What are the market signals that matter to Dmitry's decision of joining a specific community, and what roles and purposes do these communities or platforms play within the broader ecosystem? Ultimately, we build towards our thesis that by studying these mechanisms we could `explain', and therefore act upon, Dmitry's choice of joining a criminal community rather than another.
To build this argument, we first discuss previous work evaluating differences in trust signals depicted in criminal forums. We then present preliminary results evaluating criminal channels on Telegram using those same lenses. Further, we analyze the different roles these channels play in the criminal ecosystem. We then discuss implications for future research.



\section{Dmitry selects forums to join}

In~\cite{Campobasso2023-tq} Campobasso et al. laid a framework for evaluating trust-signaling mechanisms in underground forums. 
The framework comprises mechanisms that address adverse selection and moral hazard, well-known principles that are key to market efficiency~\cite{Herley2010-jp}. 
\cite{Campobasso2023-tq} derives 28 features of forum markets addressing issues associated with these and evaluate their distribution across more than 20 active underground forums. Campobasso et al. find that the forums chosen by criminals selling or providing high-impact technologies and service\footnote{\cite{Campobasso2023-tq} defines a `high-impact technology as one creating enough havoc to merit investigation and consideration by law enforcement agencies.} tended to have segregation or vendor vetting mechanisms in place, and admin roles were not tied to commercial activities in the forum. Interestingly, they find that forums that hosted `successful criminals' looked more alike, over the defined feature set, than forums that those criminals did not choose to join. Further, the feature set across the two different groups of forums is significantly different, suggesting that the 28 identified mechanisms are not randomly distributed across forums. In combination, these insights suggest that market participants do not randomly choose which communities to participate in: mechanisms addressing market inefficiencies do matter to Dmitry's decision to join.

\section{Dmitry selects Telegram spaces to join}

The Telegram platform is progressively becoming important in the criminal landscape~\cite{essay96173}. To better understand this emerging ecosystem, we executed two preliminary studies to evaluate the trust building mechanisms adopted by a set of Telegram criminal channels and groups, and the role that these groups have in supporting `traditional' criminal markets.

\subsection{Trust signals on Telegram}

To understand this emerging ecosystem, we first employ the framework defined by Campobasso et al. for a selection of mainly Russian-speaking Telegram channels. To do this, we collect messaging data from 543 groups and channels, for a total of approximately 1.1M messages. To collect these, we start with data from two repositories of Telegram groups and channels attributed to data leaks and cybercrime activities~\cite{TGDatasetRepo2023,FastFireTelegram2024}. We employ these data sets as seeds to discover (using the methods introduced in~\cite{tarik}) additional groups and channels related to cybercrime. We collect these data in three batches. The first two batches are used to build and check for saturation of a taxonomy of trust mechanisms adopted in groups and channels seeded from \cite{TGDatasetRepo2023}. The third batch is used to validate the taxonomy using a collection of groups and channels seeded independently from batches 1 and 2. To build the taxonomy, we manually analyzed posts in the first batch and mapped the mechanisms identified in~\cite{Campobasso2023-tq} using a card-sorting mechanism. The process was iterative and involved two authors independently coding randomly selected sub-batches of the data until the codes stabilized. The derived taxonomy was then applied to batch 2 and 3 for verification. This process was iterative, and when a code had to be updated based on new analysis, changes were applied retroactively to assure consistency. Table~\ref{tab:taxonomy} reports the final taxonomy.

We find that a small fraction of the forum mechanisms identified in~\cite{Campobasso2023-tq} apply to the Telegram ecosystem. This suggests that the environment is much less mature and may hinder its employment as a platform to trade highly technological and innovative criminal products such as new malware or criminal services infrastructures~\cite{Akerlof1970-hc,Herley2010-jp}. A preliminary exploration of the data through keyword matching shows that these mechanisms are not only few but also rarely signaled. This means that the trust signals a user receives when joining are likely to be weak and uncertain. This casts doubts on Telegram being a suitable platform to share high-tech criminal products with novel users. We, therefore, hypothetise that given an occasional user finding criminal channels by `snowballing' from well-known starting points (a dynamic our sampling mechanism replicates), Telegram may at most support low-value crime activities (e.g. trading of stolen data). 

\begin{table}[t]
\centering
\begin{tabular}{p{0.1\textwidth} p{0.2\textwidth} p{0.7\textwidth}}
\toprule
Type & Mechanism & Description\\ 
\midrule
\multirow{8}*{\rotatebox[origin=c]{90}{Payment methods}}& Escrow Service & Mechanisms to escrow money before transfers between vendors and buyers. Typically advertised as a third party service.\\
 & Telegram Bots & Automated means to purchase illegal products, removing uncertainty in the money transfer.\\
 & Digital Wallets & Digital wallets such as PayPal, CashApp, and Telegram Wallets facilitating easy and quick transactions. \\
 & Crytocurrency & Employment of virtual currencies such as Bitcoin and Tether (USDT) enabling perceived anonymous payments. \\
 \midrule
\multirow{2}*{\rotatebox[origin=c]{90}{Mod.}} & Rules & Guidelines set and enforced by administrators or by sellers and are used to regulate user behaviors and interactions. \\ 
\midrule
\multirow{6}*{\rotatebox[origin=c]{90}{User Verification}} & Vouching sys & Members of a group or channel endorse or vouch other users on their trustworthiness. \\
 & Scam reports & User warnings of fraudulent activities generally backed up by screenshots or forwarded messages. Can lead to banning. \\
 & Reviews \& feedback &  These refer to comments given by users based on their experience with a particular seller, service, or product.\\
\bottomrule
\end{tabular}
\caption{Taxonomy of trust building mechanisms in Telegram. (Mod.: Moderation).}
\label{tab:taxonomy}
\vspace{-0.3in}
\end{table}


\subsection{Emerging roles of Telegram in the cybercrime ecosystem.}

`Snowballing' from well-known Telegram places may, however, not be the only way Dmitry finds criminal spots to join. We observe that many posts on underground forums advertising criminal products do link to Telegram. This could be a more focused way to enter Telegram's criminal space, whereby Telegram spots inherit trust-signaling mechanisms from forums while centralizing activities like automated payments and customer support.

We, therefore, run a second preliminary study evaluating the role that Telegram channels and groups play about `traditional' forum-based criminal activities. We start from the \textit{Malware, Carding, Servers, Documents, \emph{and} Traffic} sections of the forums \textit{CryptBB, Cracked, CrackingPro, Nulled, 2crd, \emph{and} Xss}. We randomly sampled 500 threads advertising a product in those categories and referring to at least one Telegram channel or group in the advert. Following the same labeling method outlined in Sec 3.1, we identify six roles: \emph{Private Communication} facilitates one-on-one private communication between users; \emph{Automated Services}
 enable a variety of services without the need for human intervention. These services can include functionalities such as automated transactions or encrypting of files run by a Telegram bot; \emph{Announcements} Channels/groups share updates, information regarding products and services, discount codes, and product restocks. \emph{Proof of Credibility} provide evidence of the credibility of the vendor. This can include photographs or documents, screenshots of work, and forwarded reviews from previous customers; \emph{Marketplace} channels and groups offer a variety of products and services that can be purchased, typically including prices and descriptions; finally, \emph{Community Building} facilitates many-to-many communication, typically for discussion, collaboration, and knowledge sharing.

\begin{figure}[t]
    \centering
    \includegraphics[width=0.8\linewidth, height = 4.3cm]{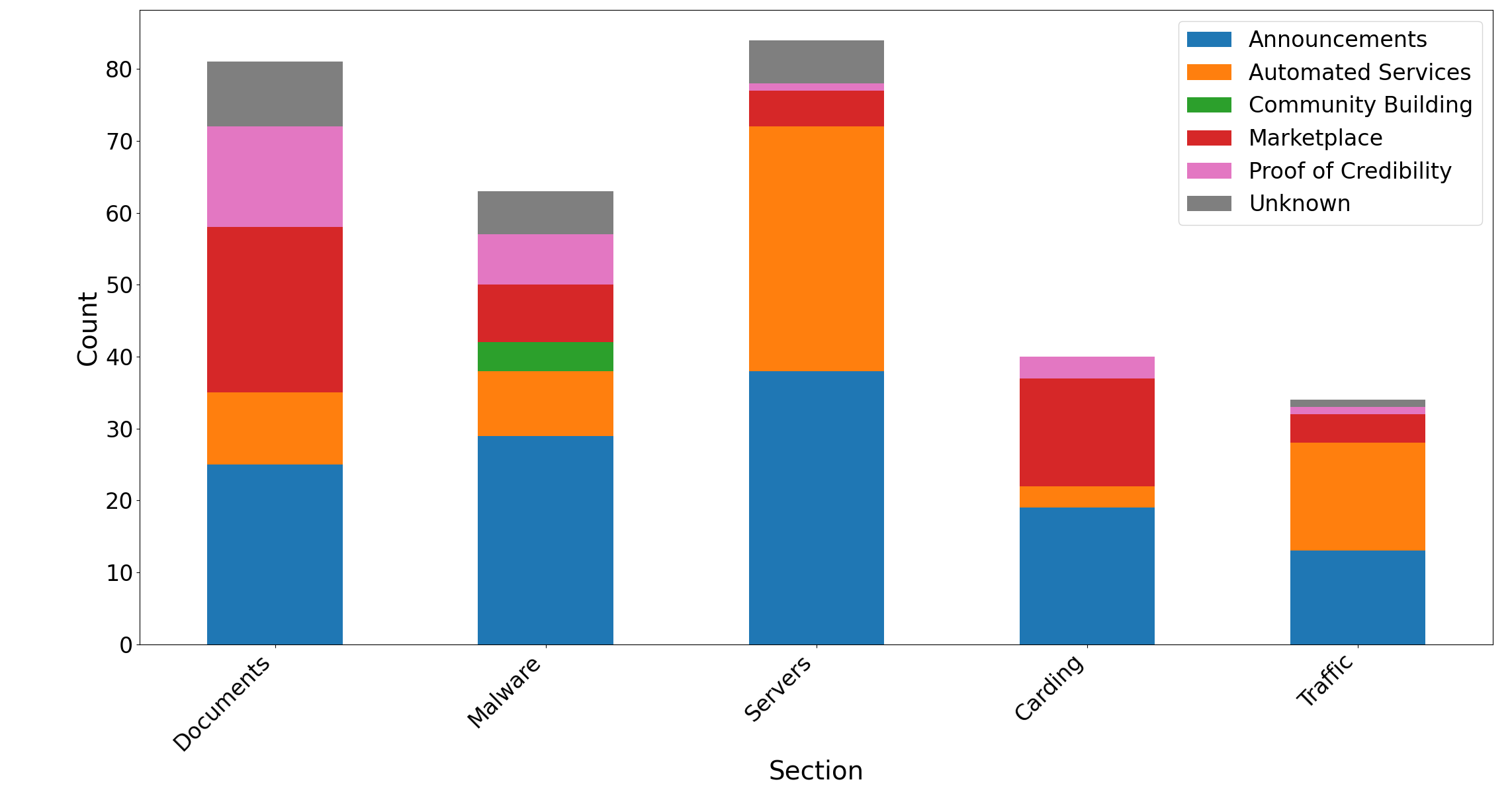}
    \caption{\textbf{Identified roles of forum-linked Telegram channels and groups.}}
    \label{fig:BarPlotFiltered}
\end{figure}

Figure~\ref{fig:BarPlotFiltered}
provides a view of role distributions across forum sections for all forums. The category `private communication' (excluded from the visualization) is by far the most used. We observe that the Marketplace role (red) is particularly common for Documents, Carding, and Malware but relatively low-prevalence for other categories. Proof of credibility is prevalent in Documents and Malware, suggesting trade there benefits from showcasing of the final products. In contrast, automated services seem to mostly support server and traffic trading. This makes sense as purchasing bulletproof services and internet traffic is a highly automatable procedure. Announcements seem prevalent across all categories. Separately, we find that some roles tend to appear together in product adverts. For example, the Announcement role is often paired with Marketplace, suggesting that the former is used to update subscribers on products sold in the latter. Similarly, automation and private communication are also often paired: the former automating service delivery, the latter providing customer support.

This suggests that the Telegram ecosystem is evolving and can effectively support traditional crime activities perpetrated through underground forums. A coherent ecosystem of services and customer relationships may be emerging on Telegram, not suffering from an otherwise likely fatal lack of mechanisms to control trade~\cite{Akerlof1970-hc}. 

\section{Way forward and conclusions}




These preliminary results provide evidence that, whereas trust building mechanisms do seem to affect Dmitry's decision to join a form rather than another~\cite{Campobasso2023-tq}, those mechanisms appear to be only weakly implemented in emerging Telegram channels. This casts doubts on the relevance of Telegram in the overall threat landscape, particularly on key questions such as threat innovation and threat intelligence. 
Nonetheless, some Telegram groups and channels seem to be forming a coherent ecosystem whereby a composition of Telegram services supports the trade and provision of advertised products on criminal forums.

We consider this initial evidence that not all Telegram places are the same, and that the emerging ecosystem within which Dmitry moves is complex. 
Although a comprehensive model of such movements is far away (if possible at all), the underlying mechanisms may have interesting parallels with factors impacting the economics of migration. \cite{czaika2021migration} For example, migrants often consider data on job markets and living conditions. News, social media, forums, and related community network effects also impact perceptions of migration risks and opportunities. These mechanisms could similarly influence Dmitry's decisions, swayed by community reputation, emerging economic opportunities, or law enforcement presence.
Naturally, particularly in the underground, these decisions are fraught with incomplete information, misinformation, and bounded rationality. These may therefore be better thought of as `prospects' whereby decisions are governed by perceptions of gains and losses relative to the current status quo. These have been considered in the migration literature~\cite{czaika2015migration}, and may offer valuable insights on how `Dmitry' positions himself in the underground too. Whereas modeling single actors is unreasonable, we speculate that thinking in terms of factors attracting specific users towards, or pushing them away from, underground communities signaling certain properties may ultimately lead to answering the question: 

\smallskip
\noindent\textit{Which `type' of Dmitry will position themselves in which `type' of community, and how can this inform threat intelligence and law enforcement operations countering the threat they generate?}





%
%
%
\bibliographystyle{splncs04}
\bibliography{References}
%




\end{document}